\documentclass[manuscript,review=false]{acmart}
\AtBeginDocument{%
  \providecommand\BibTeX{{%
    \normalfont B\kern-0.5em{\scshape i\kern-0.25em b}\kern-0.8em\TeX}}}

\copyrightyear{2023}
\acmYear{2023}
\setcopyright{rightsretained}
\acmConference[IDC '23]{Interaction Design and Children}{June 19--23, 2023}{Chicago, IL, USA}
\acmBooktitle{Interaction Design and Children (IDC '23), June 19--23, 2023, Chicago, IL, USA}
\acmDOI{10.1145/3585088.3593862}
\acmISBN{979-8-4007-0131-3/23/06}

\usepackage{tabularx,booktabs}

\usepackage{caption}
\usepackage{subcaption}

\begin{document}

\title{Designing Empathy Game: Case on Participatory Design Session with children within the Indian context
}


\author{Ekaterina Muravevskaia}
\email{}
\affiliation{%
  \institution{AMMACHI labs, Amrita Vishwa Vidyapeetham}
  \streetaddress{}
  \city{Kollam}
  \state{Kerala}
  \country{India}
  \postcode{690525}
}
\email{ekaterina.muravevskaia@ammachilabs.org}

\author{Abhijith A}
\affiliation{%
  \institution{School of Social and Behavioral Science, Amrita Vishwa Vidyapeetham}
  \streetaddress{}
  \city{Kollam}
  \state{Kerala}
  \country{India}
  \postcode{690525}
}
\email{amhsp2scw20014@am.students.amrita.edu}

\author{Pranav Prabha}
\affiliation{%
  \institution{AMMACHI labs, Amrita Vishwa Vidyapeetham}
  \streetaddress{}
  \city{Kollam}
  \state{Kerala}
  \country{India}
  \postcode{690525}
}
\orcid{0000-0002-8547-5691}
\email{pranav.prabha@ammachilabs.org}

\author{Mahesh S Unnithan}
\affiliation{%
  \institution{AMMACHI labs, Amrita Vishwa Vidyapeetham}
  \streetaddress{}
  \city{Kollam}
  \state{Kerala}
  \country{India}
  \postcode{690525}
}
\email{maheshs.unnithan@ammachilabs.org}

\author{Arunav H}
\affiliation{%
  \institution{Department of Mechanical Engineering, Amrita Vishwa Vidyapeetham}
  \streetaddress{}
  \city{Kollam}
  \state{Kerala}
  \country{India}
  \postcode{690525}
}
\email{am.en.u4are22011@am.students.amrita.edu}

\author{Rthuraj P R}
\affiliation{%
  \institution{Department of Mechanical Engineering, Amrita Vishwa Vidyapeetham}
  \streetaddress{}
  \city{Kollam}
  \state{Kerala}
  \country{India}
  \postcode{690525}
}
\email{am.en.u4are22032@am.students.amrita.edu}

\author{Ruthvik Kanukunta}
\affiliation{%
  \institution{Department of Computer Science Engineering, Amrita Vishwa Vidyapeetham}
  \streetaddress{}
  \city{Kollam}
  \state{Kerala}
  \country{India}
  \postcode{690525}
}
\email{am.en.u4cse20062@am.students.amrita.edu}

\author{Gayathri Manikutty}
\affiliation{%
  \institution{AMMACHI labs, Amrita Vishwa Vidyapeetham}
  \streetaddress{}
  \city{Kollam}
  \state{Kerala}
  \country{India}
  \postcode{690525}
}
\orcid{0000-0003-2245-1550}
\email{gayathri.manikutty@ammachilabs.org}

\renewcommand{\shortauthors}{Muravevskaia, et al.}

\begin{abstract}
Empathy games are a promising yet new research avenue that explores how to design empathic game experiences that would help children to understand and address the emotions of other people. Research in this field was primarily done in the USA and there is a research gap in understanding how empathy game design can apply and differ from the contexts of other countries. Our study replicated a study earlier conducted in the USA, aiming to explore the dynamic of the PD process, and identify specifics and challenges for PD methodology related to empathy and game design in the Indian context. We conducted a series of participatory design (PD) sessions with 18 Indian children between 7 and 11 years old. This paper reports our preliminary findings, including the following: (1) it might be challenging for Indian children to discuss and design for empathy and emotions-related topics, (2) using the English language can be a barrier while working with Indian children of 8 years old and younger, (3) cultural context affects roles children play in the design process. This paper contributes insights on identifying areas for further methodological work in PD for the Indian context.

\end{abstract}





\maketitle

\section{Introduction and Background}
Social Emotional Learning (SEL) is crucial for young children as it helps them cultivate pro-social behaviours. One of the components of social-emotional learning (SEL) in elementary school children is empathy development, an essential yet challenging educational process. Although SEL is gaining popularity in the west, it is still in its early stages in India. This is particularly true for rural and semi-urban schools in India, where structured SEL programs still need to be widely supported by the educational system. \cite{Khazanchi2021}. For teaching SEL, several studies have indicated that games are a promising medium for promoting empathic experiences (\cite{schrier2021systematic}, \cite{belman2010designing}, \cite{Greitemeyer2010}). Recently, Muravevskaia et al. \cite{muravevskaia2023designing} explored the use of empathy games to address developmental challenges related to empathy in children, such as heightened bullying and social isolation during the COVID-19 pandemic. However, there is a lack of such studies on empathy games in the Indian context.

When it comes to designing games for children, a participatory design approach is an excellent method for engaging children in the ideation process (\cite{Khaled2014}, \cite{muravevskaia2023designing}). It facilitates mutual learning and enables adult design partners to view problems from a child’s perspective, providing a better understanding of the challenges they may face. A study conducted by Sharma et al. \cite{Sharma2022} on neurodiverse individuals from special schools in New Delhi, comprising of children aged 10-18 years and adults over 18 years, emphasized the significance of cultural translation and including a local language speaker as translator or researcher due to potential language barriers. In addition, the research recommends including teachers or parents in the design process or finding ways to engage familiar adults with children. In another study \cite{Kam2006} conducted at a rural primary school in northern India with ten girls and two boys aged 10-16 years old, researchers found that an ice-breaker activity before each session was crucial in enhancing child-adult interactions. They also noticed that the children's excitement at being videotaped wore off quickly, and the participants did not react nervously to video cameras after the initial few minutes. These studies encourage further exploration into children's role as design partners through a cultural translation lens when they co-design games with professional game designers.

 Our paper describes a study on designing empathy games with Indian children. It replicates an earlier participatory design study conducted with children in the US in 2016 \cite{muravevskaia2023designing}. While being interested in participatory design in general, this study concerns a specific design method that has been shown to work in Child-Computer Interaction community before \cite{muravevskaia2023designing}. By conducting this replication study, we aimed to examine how the participatory design format with a focus on empathy and game design is perceived by children. In particular, we were looking for insights on the children's contributions to the design process, the roles they assume during the activity, how their social context and background may influence the session's outcomes, and how researchers' behavior can affect the participatory design process. This paper is organized as follows. In Section 2, we discuss our study method, data collection, and analysis. In section 3, we present our findings. We conclude with a discussion of our findings in Section 4.

\section{Methodology}

\subsection{Procedure and Apparatus}

\begin{figure}[b]
\centering
\begin{subfigure}{.5\textwidth}
  \centering
  \includegraphics[width=.8\linewidth]{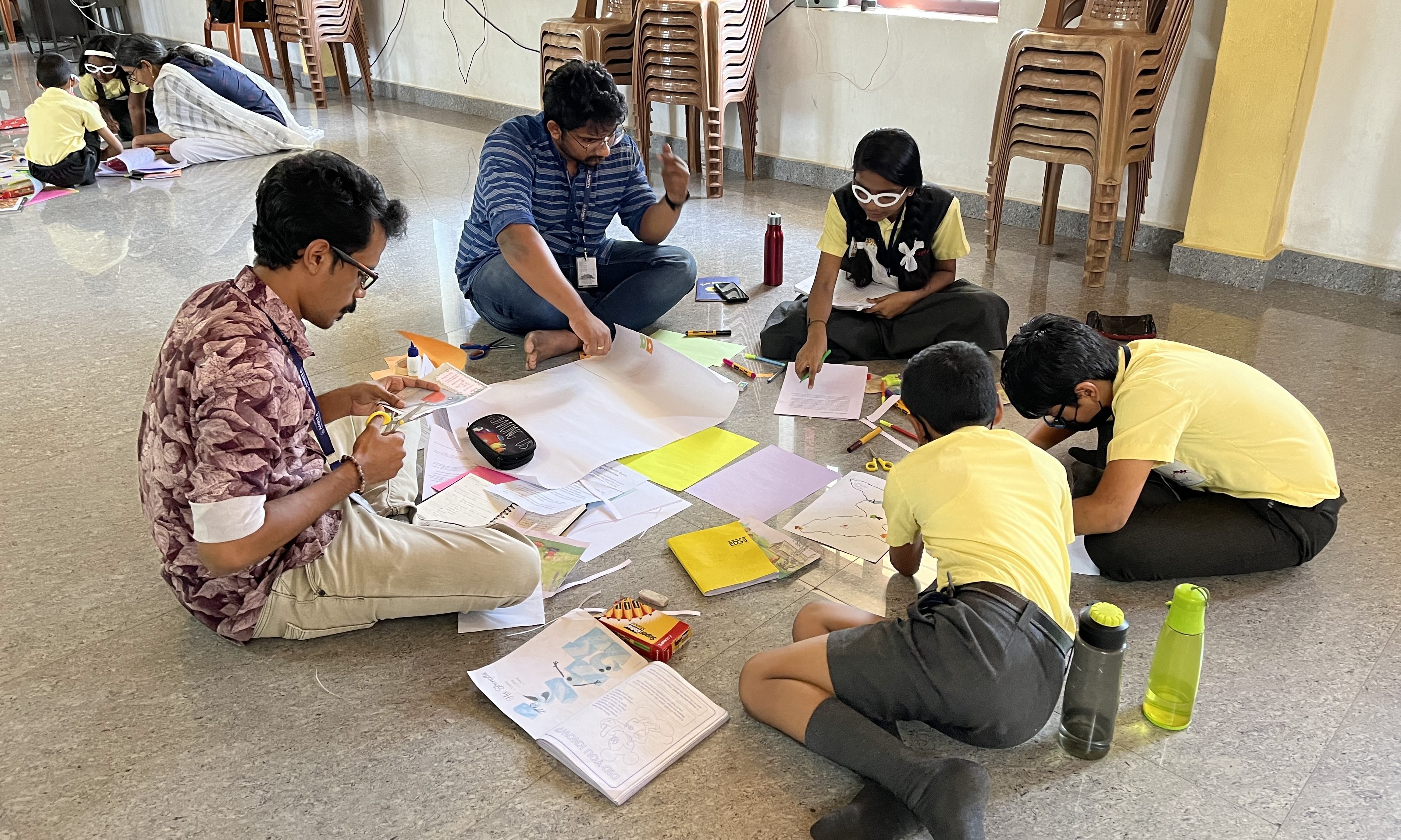}
  \caption{Participatory Design Session}
  \label{fig:sub1}
\end{subfigure}%
\begin{subfigure}{.5\textwidth}
  \centering
  \includegraphics[width=.8\linewidth]{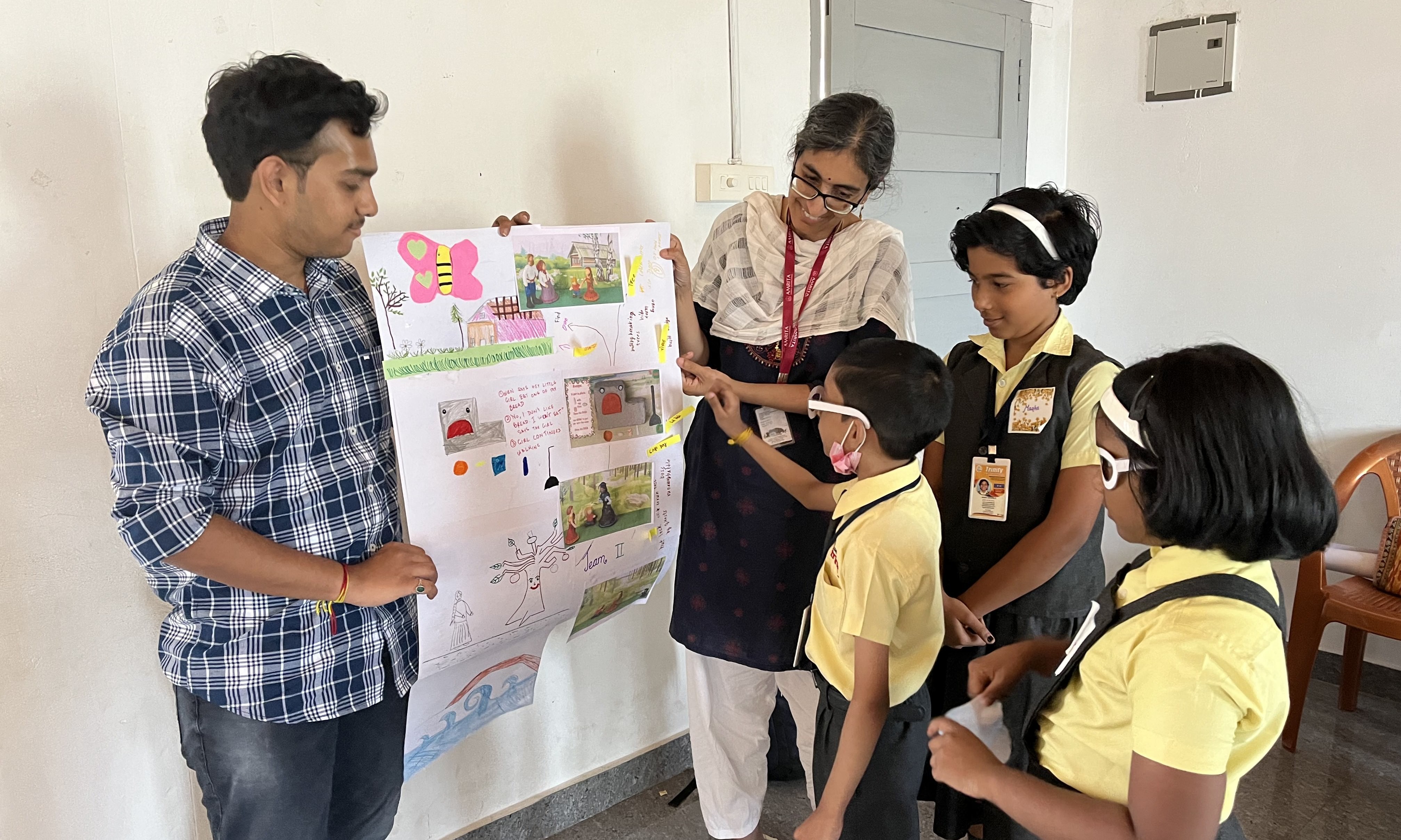}
  \caption{Presentation}
  \label{fig:sub2}
\end{subfigure}
\caption{Different phases of the study}
\label{fig:test}
\end{figure}

We based the study on the “Five Elements” and “Game Motif” participatory design methods suggested by Uğraş et al.\cite{uugracs2022new} for game design.This study took 120 minutes including a 90 minutes Participatory Design (PD) session. Each session included four main phases: 1) Circle time, 2) Participatory Design Session, 3) Presentation, and 4) Member Checking.

\textbf{Circle time}: We started with an ice-breaking activity followed by discussion on how one can understand others' feelings and why it is important. We explained the rules and goals for the design session (i.e., to design a game helping a player to understand how the characters feel). \textbf{Participatory Design}: The session included ideation and designing the game using the Big Paper prototyping technique \cite{walsh2013facit}. We formed three teams each with three children and two adult design partners. Each team received materials (i.e., chart paper, colored paper, pencils, scissors, glue, markers), a text of the fairytale "The Magic Swan Geese" as a storyline foundation, printouts of the fairytale's characters, and paper glasses representing characters’ perspectives. \textbf{Presentations}: After the design process was finished, each team presented their game ideas to others. One of the researchers was making notes on the key points of the presentations of each team. \textbf{Member Checking}: Then, we briefly discussed the key points that we had noted with the children to ensure the correctness and accuracy of note-taking.

\subsection{ Participants}
Eighteen children (7-11 years old) and six adults (20-49 years old) participated in our study. Children were selected by the school teacher based on the following criteria: confident, extroverted, and interactive.  The children were randomly split into two groups of nine. The first group (A) attended the study on Day 1 and Day 3. The second group (B) attended the study on Day 2. Although the split was random, the group A had primarily older children (11, 10, and 9 years old) with only one 8 and one 7-year-old. The group B had mostly younger children (7 and 8 years old with only one 9 and three 10-year-olds). 

\begin{table}[t]
\label{table: Program Schedule}
\begin{tabularx}{0.85\textwidth}{@{}c*{10}{c}c@{}}
Day    &Group              &Circle Time                &Participatory Design           &Presentation        &Member Checking  \\\midrule
1       &A              &12:55 - 1:55 PM        &1:55 - 2:35 PM     &2:35 - 2:55 PM       &2:55 - 3:00 PM  \\
2       &B              &11:00 AM  - 12:00 PM      &1:00 - 2:30 PM     &2:30 - 2:55 PM       &2:55 - 3:00 PM    \\
3       &A                     &N/A             &1:30 - 2:30 PM     &2:30 - 2:55 PM       &2:55 - 3:00 PM    \\
\end{tabularx}
\vspace{2mm}
\caption{Program Schedule}
\vspace{-7mm}
\end{table}

\subsection{ Data Collection }
This study was approved by the ethics committee IEC.AIM 5.2023.ANIMACHI-41. Parental consent was collected in children's native language (Malayalam) and in English. We conducted this study mainly in English. Even though it is not the native language of the children, it is the main language of the school's educational process. Prior to the study, we had a brief unstructured interview with a teacher about children’s teamwork experience, creative capacity, and communication skills. Children and adults were divided into three teams (3 children and 2 adults). The study session started by getting verbal assent from each child. For anonymity purposes, children wrote their favorite character on their nametags. During the study, we collected observation notes, debrief audios, videos, pictures, and game artifacts created during PD sessions. The initial plan was to conduct the study in two days as 150-minute sessions each. In contrast to the earlier study with American children, Indian children were not familiar with the PD process, so we extended the original study time by 30 minutes to introduce  children to the PD procedures. However, the study took more time than expected. In particular, on Day 1, due to logistical problems, we started 25 minutes later. The ice-breaking session was also extended from 30 minutes to 60 minutes. See Table 1 for the timing details of each day. As children had to leave by 3 05, the PD session time was shortened to 35 minutes. After the first day of study, we made adjustments for Day 2 (Group B) and decided to compensate the PD time for Group A and invited them to participate on Day 3 to make sure they completed the game design process properly. In addition, younger children had challenges understanding instructions in English. Therefore, we needed to translate instructions and conduct PD sessions in the children's native language. We qualitatively analyzed the data using the reflexive thematic analysis method with five researchers \cite{nowell2017thematic}, inductive coding, and thematic analysis. To establish internal validity, we conducted member checking with children as the final part of the study. 

\section{ Findings}

All the children who participated in the study were engaged throughout the entire session. They found the activity enjoyable and "very fun." Most of the children asked researchers to conduct a similar design activity again. Based on our observations and debriefing notes, we identified the following preliminary themes, which offer insights into how the children approached the participatory design process.


\textbf{Theme 1: Challenges with Designing for Emotions-Related Context}

During the design session, children seemed uninterested in generating ideas related to emotions and empathy and were more focused on the game mechanics. Despite the reminders about the study objective by the adult design partners and the follow-up questions such as "how does your idea help a player to understand characters' feelings?", children did not develop their game ideas towards characters' emotions. Only after repeated prompting from adult design partners did children generate a few ideas. Their presentation also lacked evidence of game design ideas promoting emotions and empathy besides two instances (i.e., "empathy weapon" and "golden glasses"). However, during circle time, the majority of  children were able to contribute to discussions on understanding others' feelings and the importance thereof (e.g., looking at others' faces, asking and listening, and engaging in activities together).

\textbf{Theme 2: Children's Game Design Ideas \& Age Dynamics}

During the ice-breaking activity, while discussing the favourite games of the children, we noticed that: (1) children from Group A (older children) mentioned video games as their favorite games, and children from Group B (younger children) more often referred to physical games; (2) game design projects of the children from Group A were more influenced by the video games they played (e.g., cut scene, coins, main boss fight, chasing the villain, mini-games levels, the last level being the hardest level, a linear plot of the game narrative following the storyline). Even though the older children developed more advanced features/controllers, flow, and rules of the game, we observed that they mainly incorporated game elements from popular video games. On the other hand, the younger children (Group B) added more creative alterations to the game narrative and aesthetics (i.e., making alterations to the storyline, adding additional or altered characters, adding alternative game narrative branches, turning the game into another game such as a hide-and-seek game). In addition, we observed teamwork dynamics among children and noticed that some older children motivated younger ones to contribute their creative ideas.

\textbf{Theme 3: Language Barriers for Children of 8 Years Old and Younger}

Children of 9 years and older spoke English language fluently (including reading, writing, speaking, and listening) and did not need additional language adjustments. Younger children (7-8 years old), however, were only able to write and read in English. They were more comfortable speaking and listening in their native language (i.e., Malayalam). For example, when we asked them whether they understood what the English-speaking researcher said, they said that they did. However, we noticed there was confusion on their faces. Therefore, in order to confirm whether they understood or not, a researcher who speaks the Malayalam language asked children whether they needed a translation of what was being said in English. Most of the children nodded yes. In addition, during the PD session with younger children, communication gradually shifted to Malayalam. With mixed teams including older and younger children, older children were speaking English, and the adult design partner needed to translate into Malayalam for the younger children.

\textbf{Theme 4: Importance of Ice-Breaking before the PD Session }

To create a more comfortable design environment for the children and help them to build a bond with the adult design partners, we began the study with a 15-minute ice-breaking activity. It helped to create a friendly and open atmosphere for all the participants and a smooth transition to the design process. On the second day of the study, one of the researchers missed the ice-breaking activity and joined directly into the PD process. We observed that the children who worked with this researcher were more reserved and shy at the beginning of the PD session. It took them about 15 minutes to feel comfortable enough to actively engage in the design process.

\textbf{Theme 5: "Yes Ma'am" Effect: Child-Adult Interaction within the Cultural Context}

Prior to the study, one of the adult design partners from India shared her observations (based on her prior work with school children in India) about hierarchy within the Indian classroom and how it affects the ways children communicate with adults.  We have noted already how children nodded they understood instructions in English, but actually needed the instructions in their own language. This highlights the importance of checking whether children actually understand, even if they say they do. Another example of power differential was seen when a researcher stood up and wrote notes on the blackboard (instead of sitting in the circle with the children) which elicited a more formal and distant behavior from the children.

\section{Discussion}

Our preliminary findings contribute to the interaction design community by providing insights on how to adapt the participatory methods to the Indian context. In particular, this paper identifies areas for further methodological work in PD for the Indian context including 1) potential language barriers and different age dynamics, 2) cultural context nuances, and 3) challenges with social-emotional learning.


Based on observations and debriefing notes, we suggest that in preparing for a PD session in an Indian context, take into account children's needs and comfort level related to language, familiarity with adults, and environmental settings. For example, for younger children (under 9 years old) researchers should prepare study scripts in children's native language as well as English. In agreement with Sharma et al.\cite{Sharma2022}, who worked with children between 10 and 18 as well as adults, we suggest that it is beneficial to have translators present or researchers who are comfortable speaking the children's native language. However, in contrast to Sharma et al.\cite{Sharma2022} who suggested including teachers or parents in the design process, based on our observations, we did not find it necessary. Instead, we suggest including an ice-breaking activity before the study. In agreement with Kam et al., \cite{Kam2006}, who worked with older children aged 10-16, we found that an ice-breaker at the beginning of each session is important for younger children (7-9 years old) as well. However, in contrast to Kam et al.\cite{Kam2006} findings, we found no excitement or interest in children with video cameras. We suggest that the way children approach the study equipment may vary depending on children's prior exposure to technology.  Researchers should consider testing equipment with a particular audience during the pilot study to ensure its suitability.


Participatory design can be effective for engaging Indian children in design activities. However, the traditional instruction-based teaching approaches and power dynamics between children and adults (\cite{sriprakash2010child},\cite{Clarke2003}) might make Indian children more sensitive to adults’ requests and responses \cite{Joshi1994}. This can complicate the goal of equal partnership between adults and children during the PD process. Therefore, we suggest that researchers be more conscious of their behavior and attempt to interact with children on their level to establish a more collaborative partnership. This could involve sitting on the floor together, avoiding authoritative instructions, and other actions that reduce the distance between child and adult partners. Researchers need to be aware that younger children may tend to say "yes ma'am" without meaning it and repeat others' answers during ice-breaking and brainstorming activities. To ensure all children are engaged and thinking for themselves, researchers need to pay special attention and be creative in their approach. One potential solution is to avoid popcorn-style discussions and, instead, have children write their answers first, then read them out loud or act them out. Alternatively, researchers could ask children to explain what they meant in their own words, rather than simply repeating what others have said. Further work needs to be done to explore how “equal partnership” in PD between adults and children \cite{Yip2017}, and co-design group dynamics (\cite{van2014exploring}, \cite{van2015challenging}) might work for the Indian context. 


Researchers must be mindful of how they discuss empathy and emotions-related topics with Indian children. There is to be explored which tools and methods are best suited for helping children better connect with these concepts. It may be beneficial to provide additional sessions on basic social-emotional learning (SEL) concepts and provide more details for children at the outset of each session. Future studies could investigate how to conduct PD sessions that are tailored to the specific domain of SEL, while also exploring how to effectively combine domain expertise and procedural knowledge within the context of the Indian culture \cite{Khaled2014}.

\subsection{Limitations}

Logistical challenges caused 55 minutes delays during the first day of the study. Therefore, we invited Group A to finish their project on the third day to ensure children from Group A had sufficient time to complete their game ideas. These unanticipated changes may have caused slight variations in the children's experience of the PD process for Group A and Group B. Also, the principal researcher, who was the only Caucasian on the team, received more attention and respect from the children. But it's unclear if it was due to her race or other factors. The school teacher did not think the researcher's race affected the study. 

\subsection{Future work}

 We suggest follow-up studies on (1) what roles single-age and mixed-age groups children take in the PD process and whether power differentials equally affect older and younger children, (2) what ways of PD methodology can be helpful to design for emotional contexts and empathy games. In a broader perspective, our work has considered a part of a design process at an operational level, but it is worth, as Sharma et al. \cite{Sharma2022} suggested to take a broader perspective and consider issues of local contextualized political and cultural practices and epistemologies, so how does the value system from Scandinavia needs to be translated to the local context. Our next steps in this research will include analysis of the collected data (1) to gather children's thoughts and perspectives on the flow and functionality of an empathy game and (2) conduct comparative data analysis on PD sessions between USA and Indian contexts. Then, we plan to conduct mixed methods follow up studies that looks onto a full design process including game design and development.

 \section{Selection and Participation of Children}

 This study was approved by the ethics committee IEC.AIM 5.2023.ANIMACHI-41. Before the study, parental consent was collected in the children's native language (Malayalam) and in English. We conducted the study with 18 child design participants (7-11 years old) and six adult design participants (20-49 years old). Children were selected by the school teacher based on the following criteria: confident, extroverted, and interactive.  The children were randomly split into two groups of nine children each - the first group, group A, attended the study on Day 1 and Day 3. The second group, group B, attended the study on Day 2. Although the split was random, the first group had primarily older children (11, 10, and 9 years old) with only one 8-year-old and one 7-year-old. The second group had mostly younger children (7 and 8 years old with only one 9-year-old and three 10-year-olds). The study session started by getting verbal assent from each child to participate in the study. For anonymity purposes, children were asked to write on their nametags the name of their favorite character. During the study, we collected observation notes, debrief audio recordings and transcripts, video recordings, pictures, and artifacts created during participatory design sessions.

\bibliographystyle{ACM-Reference-Format}
\bibliography{idc}
\end{document}